\documentclass[twoside,reqno]{HERON}
\usepackage{epsfig,cite,colordvi}
\usepackage{graphicx}
\usepackage{url}
\usepackage{amssymb,amsmath,amscd,epsf}
\usepackage{times}
\usepackage{makeidx}
\usepackage{hyperref}
\usepackage[utf8]{inputenc}
\usepackage{here}
\usepackage{epstopdf}
\usepackage{subcaption} 

\begin{document}

\title{Comparaison between Coulomb and Hulthèn potentials within Bohr Hamiltonian for $\gamma$-rigid nuclei in the presence of  minimal length}

\runningheads{Preparation of Papers for Heron Press Science Series
Books}{ M. Chabab, A. El Batoul, M. Hamzavi, A. Lahbas,
 I.Moumene, M. Oulne }

\begin{start}

 \coauthor{M. Chabab}{1},\coauthor{ A. El Batoul}{1},\coauthor{ M. Hamzavi}{2},\coauthor{ A. Lahbas}{1},\author{ I. Moumene}{1},
\coauthor{M. Oulne}{1}

\address{High Energy Physics and Astrophysics Laboratory, Faculty of Science Semlalia, Cadi Ayyad University, P.O.B. 2390, Marrakesh,Morocco}{1}

\address{Department of Physics, University of Zanjan, P.O. Box 45195-313, Zanjan, Iran}{2}

\begin{Abstract}
	In this work we solve the Schrödinger equation for Bohr Hamiltonian with  Coulomb and  Hulthén potentials within the formalism of minimal length    in order to obtain  analytical expressions for the energy eigenvalues and eigenfunctions  by means  of asymptotic iteration method. 
The obtained formulas of the energy spectrum  and wave functions  , are used  to calculate excitation energies and transition rates  of $\gamma$ -rigid nuclei  and compared with the experimental data at the shape phase critical point X(3) in nuclei.
\end{Abstract}
\end{start}
\section{Introduction}
Several  analytical solutions of the Bohr Hamiltonian with different model potentials  have been proposed.  On the other hand,  this problem is related to the evolution of  Critical  Point Symmetries concept. For example, the symmetry E (5) \cite{E5} describes the second-order phase transition between spherical and $\gamma$  -unstable nuclei, while the transition from vibratory to axially symmetric nuclei is described by symmetry X (5) \cite{X5} and X(3) \cite{Bonatsos:2005mq} which  is a special case of this latter in which $\gamma$   is fixed to    $\gamma$=0. This model has been developed with the introduction of the concept of minimal length  \cite{ML-chabab}. In this context, different model potentials  have been used such as infinite {\tiny } Square Well (ISW) \cite{Bonatsos:2005df},  the harmonic oscillator  \cite{Ermamatov:2011zz},  the sextic potential  \cite{Budaca:2016wbq}  and the Davidson one within X(3) symmetry.

In the present work  we   focused on the study of the Bohr Hamiltonian in the presence of a  minimal  length  in X(3) model with two known potentials, namely: Hulthén and coulomb, where we have obtained the expressions of eigenvalues and wave functions by  means of the asymptotic iteration method (AIM)\cite{AIM03,AIM07}. Such a   useful  method  is efficient to solve many similar problems  \cite{Chabab:2018bls,chabab2016closed}.
\section{Formulation of the Model \label{sec1}}
The Bohr Hamiltonian in the  presence of a minimal length is given by \cite{ML-chabab}
\begin{equation}
	H=-\frac{\hbar^2}{2B_{m}}\Delta+\frac{\alpha \hbar^4}{2B_{m}}\Delta^2+V(\beta)
\end{equation}
with
\begin{equation}
	\Delta=\frac{1}{\beta^2}\frac{\partial}{\partial\beta} \beta^2\frac{\partial}{\partial\beta}+\frac{1}{3\beta^2}\Delta_{\Omega}
\end{equation}
where $ \Delta_{\Omega} $  is the angular part of the Laplace operator
\begin{equation}
	\Delta_{\Omega}=\frac{1}{sin\theta}\frac{\partial}{\partial\theta} sin\theta\frac{\partial}{\partial\theta}+\frac{1}{sin\theta^2}\frac{\partial^2}{\partial\phi^2}
\end{equation}

The corresponding deformed Schrödinger equation to the first order in $\alpha$ reads as

\begin{equation}
	\left[-\frac{\hbar^2}{2B_{m}}\Delta+\frac{\alpha \hbar^4}{2B_{m}}\Delta^2+V(\beta)-E \right ] \psi(\beta,\theta,\phi) =0
\end{equation}

By introducing an auxiliary wave function
\begin{equation}
	\psi(\beta,\theta,\phi)=\left[1-2\alpha\hbar^2 \Delta\right ] \phi(\beta,\theta,\phi )
\end{equation}
we obtain the following differential equation satisfied by  $\phi $
\begin{equation}
	\left[(1+4B _{m}\alpha(E-V(\beta)))\Delta    + \frac{2B _{m}}{\hbar^2} (E-V(\beta))\right] \phi(\beta,\theta,\phi )=0
	\label{sh}
\end{equation}

By considering  the wave function as \\
$$ \phi(\beta,\theta,\phi )=\xi(\beta)Y _{L M}(\theta,\phi)$$\\
and \\
$$\Delta_{\Omega}Y _{L M}(\theta,\phi)=-L(L+1)Y _{L M}(\theta,\phi)$$
Eq\eqref{sh} transforms into \cite{ML-chabab}:
\begin{equation}
	\frac{1}{\beta^2}\frac{\partial}{\partial\beta} \beta^2\frac{\partial}{\partial\beta} \xi(\beta)+\left(\dfrac{-\Lambda}{\beta^{2}}+\dfrac{2 B _{m}}{\hbar^{2}}((E-V(\beta))-4B \alpha(E-V(\beta))^{2} )\right)\xi(\beta),\label{prin}
	\end{equation}

where $V(\beta)$  is :

\begin{itemize}
	\item The Coulomb potential:
\end{itemize}
\begin{equation}
	V(\beta)=\frac{c}{\beta} ;
	\label{coulom}
\end{equation}

\begin{itemize}
	\item The Hulthén potential :
\end{itemize}
\begin{equation}
	V(\beta)=\frac{e^{-\delta \beta}}{e^{-\delta \beta}-1} .
	\label{hulth}
\end{equation}

\section{Energy  Spectrum  }
\subsection{Hulthén potential }
Using  the new  variable $y={e^{-\delta \beta}}$, Eq \eqref{prin} becomes
 \begin{equation}
  \begin{split}
      \frac{d^{2}}{dy^{2}}\xi(y)+\frac{1}{y}\frac{d}{dy}\xi(y)&+\frac{1}{\delta^{2} y^{2}}(-\Lambda \delta^{2} \frac{y}{^{(y-1)^{2}}}\\
	 +&\frac{2 B _{m}}{\hbar^{2}}((E-\frac{y}{^{(y-1)}}-4 B \alpha(E-\frac{y}{^{(y-1)}})^{2})\xi(y).
	 	\end{split}
\end{equation}

In order to apply AIM, we consider the following ansatz:
\begin{equation}
	\xi(y)=y^{\nu} (1-y)^{\mu} \chi(y)
\end{equation}
with  
$$ \nu=\dfrac{\sqrt{-2E}}{\delta}  \quad and \quad 
 \mu=\dfrac{1}{2}\left(1+\sqrt{4\Lambda+1 +\dfrac{32 \alpha}{\delta^{2}}}\right).$$\\

Using the AIM, we obtain the energy spectrum  in the following  form:
\begin{equation}
	E=-\frac{1}{8} \left[\dfrac{-\delta^{2} (\mu+ n_{\beta} )^{2}+8 \alpha+1}{\delta (\mu+n_{\beta})} \right]^{2}.
	\label{enerhul}
\end{equation}
\subsection{Coulomb  potential }
By substituting  the folowing ansatz \quad $\xi(\beta)=\beta^{\mu}e^{\nu \beta} \xi(\beta)$\quad in Eq \eqref{prin}, we get \\
\begin{equation}
\small
	\dfrac{d^{2}}{d\beta^{2}}\xi(\beta)+\left[\dfrac{2\mu+2\nu \beta+2}{\beta}\right]\dfrac{d}{d\beta}\xi(\beta)+\left[\dfrac{16   \alpha c E_{0} +2\mu\nu-2c+2\nu}{\beta}\right]\xi(\beta)
\end{equation}
with
  $$\mu=-\dfrac{1}{2}+\dfrac{1}{2}\sqrt{32  \alpha c^{2}+4 \Lambda+1}    \qquad \text{and}  \qquad 
 \nu=-\sqrt{8 \alpha E_{0}^{2}-2E}.$$
Applying the AIM, we obtain the  energy spectrum as 
\begin{equation}
	E=\frac{-2 c^2}{G}((8 \alpha E_{0}-1)^2+4 \alpha E_{0}^2)
	\label{eqcou}
\end{equation}
with
\begin{equation*}
	G=4 n _{\beta}^2+4 n _{\beta}+2+\frac{4}{3}L(L+1)+32 \alpha c^2 + 4 (2 n _{\beta}+1) \sqrt{\frac{1}{4}+\frac{L(L+1)}{3}+ 8 \alpha  c^2}
\end{equation*}
and
\begin{equation*}
	E_{0}=\frac{-2 c^2}{\left[(2 n _{\beta}+1)+ 2  \sqrt{\frac{1}{4}+\frac{L(L+1)}{3}} \right]^2}
\end{equation*}

\section{Wave functions}
\subsection{Hulthén  }
The wave function  is  written in terms of Hypergeometric functions
\begin{equation}
	\xi=N   e^{-\delta \beta \nu} (1-e^{-\delta \beta \mu})   _{2}F_{1}[-n,2\mu+2+2\nu+n,2 \nu+1,e^{-\delta \beta}],
\end{equation}
where N is a normalization constant \cite{gradshteyn}
\begin{equation}
	N=\left[\dfrac{\mu+n}{2 \delta \nu (\nu+\mu+n)} \right]^{-0.5}\left[\dfrac{(\Gamma(2 \nu +1) \Gamma(n+1))^{2}\Gamma(2\mu+n)  }{n!\Gamma(2 \nu +n+1) \Gamma(2 \nu +2\mu+n) }\right]^{-0.5}.
\end{equation}
\subsection{Coulomb  }
The wave function in this case is written   in terms of Laguerre polynomials
\begin{equation}
	\xi(\beta)=N  e^{-\nu \beta} \beta^{\mu}KummerM[-n,2\mu+2,2 \nu \beta]
\end{equation}

with
\begin{equation}
	N=\left[\left(\dfrac{1}{2\nu}\right)^{2\mu+3} \dfrac{\Gamma(n+2\mu+2)(2n+2\mu+2)}{n!(LaguerreL[n,2\mu+1,0])^{2}}\right]^{-0.5}.
\end{equation}

\section{Transition rates B(E2)}
The general expression for the quadrupole  transition operator is \cite{wilet}
\begin{equation}
	T^{E2}_{M}=t\beta \left[D^{2*}_{M,0}(\theta_{i}) cos(\gamma)+\dfrac{1}{\sqrt{2}} \left[D^{2*}_{M,2}(\theta_{i})+D^{2*}_{M,-2}(\theta_{i}) sin(\gamma)\right]\right],
\end{equation}
where t denotes a scalar factor and $D^{2*}_{M,2}(\theta_{i})$ is the Winger functions of Euler angles.

The B(E2) transition rates are given by \cite{Bonatsos:2005mq}
\begin{equation}
B(E2,nLn_{\gamma}K\rightarrow n'L'n'_{\gamma}K )= t^{2}\langle L2L'|K,K'-K,K'\rangle^{2} I_{n,L,n',L'}^{2},
\end{equation}
where $\langle L2L'|K,K'-K,K'\rangle $ are  Clebsch-Gordan coefficients and

\begin{equation}
	I_{n,L,n',L'} =\int_{0}^{\infty}\beta \xi_{n,L}(\beta) \xi_{n',L'}(\beta)\beta^{2}d\beta.
\end{equation}

\section{Numerical results }
\subsection{Spectra of $ \gamma$-rigid  nuclei }

\par The formulas of the energy spectrum,  obtained by the equations  \ref{enerhul} and  \ref{eqcou}, are used to calculate the excitation energies of $ \gamma$-rigid  nuclei. The energy spectrum of Coulomb potential depends on two parameters ($ \alpha $ , $ c $ ), while in the Hulthén potential, it depends on ( $ \alpha $ , $ \delta $). All these parameters have been set by  fitting the   excitation energies normalized to the energy of the first excited state $ E(2^{+}_{1}) $. We evaluate the root mean square (rms) deviation between theoretical values and the\\
\begin{figure}[H]
	    \centering\includegraphics[scale=0.6]{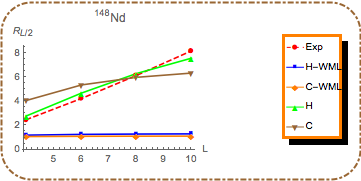}
		\centering\includegraphics[scale=0.6]{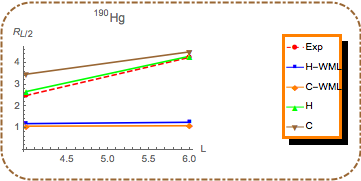}
	    \centering\includegraphics[scale=0.6]{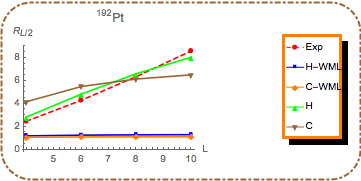}
      	\centering\includegraphics[scale=0.6]{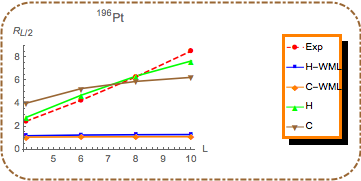}
	\caption[]{The energy  ratios in the absence of ML and in its presence}\label{f11}
\end{figure}
experimental data by \begin{equation}
\sigma=\sqrt{\frac{{\sum^{m}_{i=1}(E_{i}(exp))-E_{i}(th))^{2}}}{(m-1)E(2^{+}_{1})^{2}}}
\label{sigma}
\end{equation}
where $ m $ is the number of states, while $E_{i}(exp) $ and $ E_{i}(th) $  represent the theoretical and experimental energies of the   $i^{th}$    level, respectively.  $ E(2^{+}_{1}) $ is the energy of the first excited level of the ground state band.

   From the Eq \eqref{hulth} and  \eqref{coulom}, one can see  that both potentials 
have mathematically similar behaviors. If we give the same value to the parameter $c$ in Coulomb potential (Eq.\eqref{coulom}) and $\delta$  in the Hulthén one (Eq.\eqref{hulth}),  we  get overcome curves. 
The Figure  (\ref{f11}) shows  that    in the   absence of  minimal length  case,  the   obtained  results for energy ratios with both potentials    are identical for all  even-even nuclei, while  in its presence, the calculated energy ratios $R_{L/2}$ with Hulthén potential are
\begin{table}[H]
    \centering
    \scriptsize
    {\renewcommand{\arraystretch}{1}
    {\setlength{\tabcolsep}{0.12cm}
			\caption{{\small The comparison of  the obtained results  by the two equations: \eqref{enerhul} and  \eqref{eqcou}, for the ground  state band $(n=0)$ and the  $\beta$ band $(n=1)$ with the experimental data \cite{data}. The values of free parameters  is also shown} }
			\begin{tabular}{ c c c ccccc c c cc c}
				\hline  	\hline 
				Nucleus \qquad&$ {R_{0,4}}$ & $ {R_{0,6}}$&$ {R_{0,8}}$&$ {R_{0,10}}$&$ {R_{1,0}}$&$ {R_{1,2}}$&$ {R_{1,4}}$&$ {R_{1,6}}$&$\alpha$&$\delta$&$c$&$\sigma$\\
				\hline  
				{$^{104}$Ru}\quad{\scriptsize Exp}&2.48& 4.35&6.48&8.69&2.76&4.23& 5.81&&&&&\\
				\quad\qquad{\scriptsize H}&2.82& 4.78& 6.49&
				7.85&3.91&4.51& 5.62&&0.00003&0.003&&0.63\\
				\quad\qquad{\scriptsize C}&4.121& 5.42&6.06&
				6.42&4.11&4.63& 5.51&&0.58&&-1.31&1.36 \\
				{$^{120}$Xe}\quad{\scriptsize Exp}&2.44& 4.23&6.34&8.77&2.82 & 3.95&5.31&&&&&\\
				\quad\qquad{\scriptsize H}&2.81& 4.75& 6.43&
				7.76&3.89& 4.48& 5.58&&0.00003&0.003&&0.70\\
				\quad\qquad{\scriptsize C}&4.07& 5.35&5.98&
				6.33&4.06&4.58&5.44&&0.58&&-1.31&1.43 \\
				{$^{122}$Xe}\quad{\scriptsize Exp}&2.50 &4.43 &6.69 &9.18&3.47 & 4.51&&&&&&\\
				\quad\qquad{\scriptsize H}&2.86& 4.94& 6.81&
				8.34&4.18&4.79&&&0.00004&0.003&&0.58 \\
				\quad\qquad{\scriptsize C}&4.29&5.67& 6.35&
				6.72&4.29&4.84&&&0.58&&-1.31&1.53 \\
				{$^{124}$Xe}\quad{\scriptsize Exp}&2.48 &4.37& 6.58&8.96&3.58 & 4.60& 5.69&&&&&\\
				\quad\qquad{\scriptsize H}&2.85& 4.89& 6.71&
				8.19&4.10&4.70& 5.86&&0.00003&0.003&&0.47  \\
				\quad\qquad{\scriptsize C}&4.25& 5.61&6.28&
				6.65&4.25& 4.79& 5.71&&0.58&&-1.31 &1.32\\
				{$^{148}$Nd}\quad{\scriptsize Exp}&2.49&4.24 & 6.15& 8.19&3.04 & 3.88&5.32&7.12&&&&\\
				\quad\qquad{\scriptsize H}&2.79& 4.68& 6.30&
				7.58&3.77&4.36& 5.44&6.63&0.00003&0.003&&0.49\\
				\quad\qquad{\scriptsize C}&4.07& 5.36&5.99&
				6.34&4.07&4.58& 5.45& 6.00&0.58&&-1.31& 1.31\\
				{$^{150}$Sm}\quad{\scriptsize Exp}&2.32&3.83 & 5.50 & 7.29&2.22 &3.13 &4.34& 6.31&&&&\\
				\quad\qquad{\scriptsize H}&2.66& 4.27& 5.54&
				6.47&3.23& 3.78& 4.73& 5.70&0.00003&0.004&&0.64 \\
				\quad\qquad{\scriptsize C}&3.58& 4.66&5.19&
				5.48&3.55& 4.00&4.73& 5.19&0.58&&-1.30 &1.17\\
				{$^{152}$Gd}\quad{\scriptsize Exp}&2.19 & 3.57 & 5.07 &6.68&1.79 &2.70 & 3.72 & 4.85&&&&\\
				\quad\qquad{\scriptsize H}&2.53& 3.88&4.86& 
				5.54&2.80&3.31& 4.15& 4.93&0.00003&0.005&&0.67 \\
				\quad\qquad{\scriptsize C}&3.17& 4.08& 4.52&
				4.77&3.12& 3.52&4.14&4.53&0.41&&-1.52& 1.06\\
				{$^{172}$Os}\quad{\scriptsize Exp}&2.66& 4.63& 6.70 & 8.89&3.33 & 3.56 &5.00 & 6.81&&&&\\
				\quad\qquad{\scriptsize H}&2.81& 4.76& 6.45&
				7.79&3.88&4.47& 5.58&6.82&0.00003&0.003&& 0.63 \\
				\quad\qquad{\scriptsize C}&4.19& 5.52&6.17&
				6.54&4.18&4.72& 5.61& 6.18&0.58&&-1.31&1.29\\
				{$^{190}$Hg}\quad{\scriptsize Exp}&2.50& 4.26&&&3.07&3.77 & 4.74 &6.03&&&&\\
				\quad\qquad{\scriptsize H}&2.68& 4.31&&&3.44&3.88& 4.58& 5.02&0.00004&0.004&&0.16 \\
				\quad\qquad{\scriptsize C}&3.48& 4.51&&&3.48&4.51& 5.01&5.30&0.95&&-1.02&0.66\\
				{$^{192}$Pt}\quad{\scriptsize Exp}&2.48 & 4.31& 6.38& 8.62&3.78& 4.55&&&&&&\\
				\quad\qquad{\scriptsize H}&2.84& 4.85& 6.629&
				8.06&4.02&4.63&&&0.00003&0.003&&0.41  \\
				\quad\qquad{\scriptsize C}&4.18& 5.50&6.16&
				6.52&4.17& 4.71&&&0.58&&-1.31 &1.33 \\
				{$^{196}$Pt}\quad{\scriptsize Exp}&2.47 &4.29&6.33 & 8.56&3.19 & 3.83&&&&&&\\
				\quad\qquad{\scriptsize H}&2.80& 4.72& 6.37&
				7.68&3.82& 4.41&&&0.00002&0.003&&0.60\\
				\quad\qquad{\scriptsize C}&4.03& 5.29& 5.91&
				6.26&4.02&4.53&&&0.58&&-1.31 &1.41 \\
				\hline
			\end{tabular}
		}}
	\end{table}
		\begin{table}[H]
			 \centering
			 \scriptsize
			 {\renewcommand{\arraystretch}{1.2}
			 	{\setlength{\tabcolsep}{0.1cm}
		\caption{ 	{\small The comparison of the present model with experimental data, for theoretical predictions calculated with the two potentials: Hulthén and  Coulomb, for $B(E2) $ transition rates \cite{Bonatsos:2013oua}}}
			\begin{tabular}{ c c c c c c c c c c c}
				\hline \hline 
		 	Nucleus \qquad& $ \frac{4_{1}\rightarrow2_{1}}{2_{1}\rightarrow0_{1}} $ & $\frac{6_{1}\rightarrow4_{1}}{2_{1}\rightarrow0_{1}} $ & $\frac{8_{1}\rightarrow6_{1}}{2_{1}\rightarrow0_{1}} $ & $\frac{10_{1}\rightarrow8_{1}}{2_{1}\rightarrow0_{1}} $ &  $\frac{0_{\beta}\rightarrow2_{1}}{2_{1}\rightarrow0_{1}} $ & $\frac{2_{\beta}\rightarrow2_{1}}{2_{1}\rightarrow0_{1}}$& $\frac{2_{\beta}\rightarrow4_{1}}{2_{1}\rightarrow0_{1}} $ & 
				$\frac{2_{\beta}\rightarrow0_{\beta}}{2_{1}\rightarrow0_{1}} $ && $\sigma$\\	
				\hline
				
		     	{$^{100}$Mo} \quad Exp&1.86(11)& 2.54(38)& 3.32(49)&& 2.49(12)& 0& 0.97(49)& 0.38(11)&&\\
				\qquad \quad H&1.88&3.33& 6.16&11.26&1.52& 0.14&2.29& 2.98&&0.84\\
				\qquad\quad C&2.25& 1.41&0.76& 0.43&1.41& 4.88&0.73&0.07&&1.06\\
				{$^{108}$Ru}\quad  Exp&1.65(20)&&&&&&&&& \\
				
		       \qquad \quad H&1.59& 2.13&2.88&3.98& 0.56& 0.08& 0.58&2.04&&0.05\\
			\qquad \quad  C&2.61&1.71& 0.93& 0.53&1.98& 5.82& 0.77&0.04&&0.96 \\
				
				{$^{128}$Xe} \quad Exp&1.47(15)& 1.94(20)& 2.39(30)&&&&&&&\\
				
			\qquad \quad H&1.73& 2.65& 4.22& 6.83&0.97& 0.11& 1.25&2.52&&0.48\\
		\qquad \quad C&2.44& 1.57& 0.85& 0.48&1.71& 5.39&0.75& 0.05&&0.83 \\
				{$^{146}$Nd} \quad Exp&1.47(39)&&&&&&&&&\\
			\qquad \quad H&1.80& 2.97&5.11&8.84& 1.23& 0.13&1.73&2.76&&0.33\\
		\qquad \quad C&2.35& 1.50& 0.81& 0.45&1.57& 5.15&0.74&0.06&&0.88\\
				
				{$^{148}$Nd} \quad Exp&1.62& 1.76& 1.69&& 0.54& 0.25& 0.28&&&\\
				\qquad \quad H&1.71& 2.57&3.99&6.33& 0.90& 0.11&1.14&2.45&&0.59\\
	\qquad \quad C&2.49&1.61&0.87& 0.49& 1.78& 5.50& 0.76&0.05&&0.91\\
				{$^{150}$Sm}\quad Exp&1.93(30)& 2.63(88)& 2.98(158) & & 0.93(9)&&& 1.93&&\\
			\qquad \quad H&1.78& 2.86&4.81& 8.15& 1.14& 0.12& 1.56& 2.68&&0.57\\
			\qquad \quad C&2.40&1.53& 0.83&0.47&1.64& 5.26& 0.74&0.05&&0.76\\
				{$^{172}$Os} \quad Exp&1.56(6)& 1.82(10) &1.99(11)& 2.29(26)& 0.33(5) &0.04 &0.12(1)& 0.62(6)&&\\
			\qquad \quad H&1.70& 2.52& 3.87&6.06& 0.87&0.11& 1.07& 2.42&&1.01\\
			\qquad \quad C &2.50& 1.62&0.88& 0.50& 1.81& 5.54&0.76&0.05&&1.16\\
				
				{$^{190}$Hg}  \quad Exp&&&&&&&&&&\\
	\qquad \quad H&1.77& 2.83& 4.73& 7.97& 1.12& 0.12& 1.52& 2.66&&\\
		\qquad \quad C&2.37& 1.51& 0.82& 0.46& 1.60& 5.20& 0.74&0.062&&\\
				
				{$^{192}$Pt}\quad Exp&1.56& 1.22&&&&&&&&	\\
			\qquad \quad H&1.68&2.46&3.72&5.75& 0.82& 0.10&1.00&2.37&&0.48\\
		\qquad \quad C&2.50& 1.62&0.88&0.49&1.80& 5.54& 0.76& 0.050&&0.09\\
				
				{$^{196}$Pt}\quad Exp&1.48(2) &1.80(10)& 1.92(25)&&& &=0 &0.12&&\\
			\qquad  \quad H&1.70& 2.54& 3.93&6.19& 0.89& 0.11& 1.10& 2.43&&0.60\\
				\qquad \quad C&2.48& 1.60& 0.87&0.49& 1.77& 5.48&0.759& 0.05&&0.63\\
				
				\hline
			\end{tabular}
			\label{t2}	}}
		\end{table}
fairly better than those obtained with Coulomb one. The best candidate
nuclei  for the model with Hulthén potential are:  $^{172}$Os, $^{192}$Pt, $^{196}$Pt and $^{190}$Hg.
\section{Conclusion}
In this work, we have solved the Bohr-Mottelson Hamiltonian in the $\gamma$-rigid regime within the minimal length formalism with two well-known potentials: Coulomb and Hulthén.

From the comparison between the energy spectra and transition probabilities in the two cases: presence and absence of the minimal length, one can conclude that the obtained results with Hulthén potential within the ML are better. This latter reproduces well the $X(3) $ candidates which already have been  obtained including the predicted new one: $^{190}$Hg.

\section*{Acknowledgements}
I.Moumene would like to thank the organizing committee for giving her the greatest opportunity to attend and be one of the team members in this interesting workshop, and their hospitality during the meeting. \\
Also she would like to thank the university CADI AYYAD for the financial support. \\
A huge thank to professor  M. Oulne and  A. Antonov for their encouragement.


\bibliographystyle{elsarticle-num}
\bibliographystyle{IEEEtran}
\bibliography{bibx3}

\end{document}